\documentclass[article,twocolumn,showpacs,superscriptaddress]{revtex4-1}

\usepackage{amsmath}
\usepackage{amssymb}
\usepackage{color}
\usepackage{verbatim}
\usepackage{array}
\usepackage{graphicx}
\usepackage{epsfig}
\usepackage{bm}
\usepackage[ragged]{sidecap}
\usepackage{dsfont}
\usepackage{ulem}
\usepackage{mathrsfs}

\newcommand{\ds}{\displaystyle}

\newcommand{\supercomas}[1]{``#1''}

\newcommand{\dd}{\mathrm{d}}

\renewcommand{\vec}{\bm}

\newcommand{\opscalar}[1]{\hat{\textrm #1}}

\newcommand{\etal}{{\it et al.}} 
\newcommand{\el}{\textrm{el}}

\begin{document}

\title{
Quantized form factor shift in the presence of free electron laser radiation
}

\author{F. Fratini}
\email[]{fratini.filippo@gmail.com\\ffratini@fisica.ufmg.br}
\affiliation{Universidade Federal de Minas Gerais, Instituto de Ci\^encias Exatas, Departamento de F\'isica, 31270-901 Belo Horizonte, MG, Brasil}
\affiliation{Department of Physics, University of Oulu, Fin-90014 Oulu, Finland}
\affiliation{Institut N\'eel-CNRS, BP 166, 25 rue des Martyrs, 38042 Grenoble Cedex 9, France}

\author{L. Safari}
\affiliation{Department of Physics, University of Oulu, Fin-90014 Oulu, Finland}

\author{A. G. Hayrapetyan}
\affiliation{Physikalisches Institut, Ruprecht-Karls-Universit\"{a}t Heidelberg, D-69120 Heidelberg, Germany}

\author{K. J\"ank\"al\"a}
\affiliation{Department of Physics, University of Oulu, Fin-90014 Oulu, Finland}

\author{P. Amaro}
\affiliation{Physikalisches Institut, Ruprecht-Karls-Universit\"{a}t Heidelberg, D-69120 Heidelberg, Germany}
\affiliation{Centro de F\'{i}sica At\'{o}mica, Departamento de F\'{i}sica, Faculdade de Ci\^{e}ncias e Tecnologia, FCT, Universidade Nova de Lisboa, P-2829-516 Caparica, Portugal}

\author{J. P. Santos}
\affiliation{Centro de F\'{i}sica At\'{o}mica, Departamento de F\'{i}sica, Faculdade de Ci\^{e}ncias e Tecnologia, FCT, Universidade Nova de Lisboa, P-2829-516 Caparica, Portugal}

\date{\today}

\begin{abstract}
In electron scattering, the target form factors contribute significantly to the diffraction pattern and carry information on the target electromagnetic charge distribution. Here we show that the presence of electromagnetic radiation, as intense as currently available in Free Electron Lasers, shifts the dependence of the target form factors by a quantity that depends on the number of photons absorbed or emitted by the electron as well as on the parameters of the electromagnetic radiation. As example, we show the impact of intense ultraviolet and soft X-ray radiation on elastic electron scattering by Ne-like Argon ion and by Xenon atom. We find that the shift brought by the radiation to the form factor is in the order of some percent.
Our results may open up a new avenue to explore matter with the assistance of laser.
\end{abstract}


\maketitle

{\bf\raggedleft Introduction}

Electron scattering is a tool of great importance for exploring the structure of matter \cite{CoF2004, Seth2013, Kat2008, Gar2008, Kam2011, Wer2013}. 
The most significant example to highlight such importance is perhaps electron microscopy, where electron scattering is the core process \cite{Wil2009}. Electron scattering can be elastic or inelastic. In both cases, the diffraction pattern of the scattered electrons is influenced by the target form factor (FF), also called scattering factor \cite{Bren2008, Povh2006}. Both elastic and inelastic FFs carry information about the electromagnetic charge distribution of the target \cite{Bren2008, Povh2006}.
FFs are not only relevant in electron scattering, but play an important role also in light diffraction \cite{Chantler2000} and light absorption \cite{Ces1992}. As a consequence, FFs are subject of research in many scientific areas, such as atomic physics \cite{AlS2008}, nuclear and subnuclear physics \cite{Frosch1967, Bernauer2011}, crystallography \cite{Cry1992} and biology \cite{Tar2002, Mor1982}. 

Due to the wide applicability of FFs, exploring the effects of electromagnetic radiation (ER) on the target FFs is of general interest in science. Laser assisted electron scattering (LAES) has been the subject of intense research since the 1970s, as consequence of the development of lasers \cite{Mas1993, deHL2011, El1998, Yama2010}. In this Letter we show that the presence of intense ER in electron scattering can be used to control the FFs. More specifically, we show that the presence of ER shifts the argument of the inelastic and elastic FFs from $\vec Q$ to $\vec Q + s\hbar\vec k$, where $\vec Q$ is the momentum transfer between electron and target, $\vec k$ is the wave-vector of the ER, $\hbar$ is the reduced Planck constant, and $s$ is an integer number that represents the number of photons absorbed (if $s>0$) or emitted (if $s<0$) by the electron during the scattering process. A similar linear momentum shift on the whole Differential Cross Section (DCS) has been already highlighted in the literature \cite{EhlCanad1992, EhlOpt1988, DF1, DF2}. However, to the best of our knowledge, the linear momentum shift brought by the ER to elastic and inelastic FFs has not been clearly pointed out in the literature. In view of recent advances in X-ray Free Electron Laser (FEL) sources, it is important to highlight that exploring FF shifts has become nowadays feasible with the current state-of-the-art-technology. Quantized FF shifts can be used to tailor the information extracted from the target electromagnetic distribution. More generally, quantized FF shifts can be used broadly in science, since FFs are used in various scientific areas. As examples, in this Letter we show the effect of intense ultraviolet and soft X-ray radiation on the DCS and the FF for elastic electron scattering by Ne-like Argon and Xenon atom. Our results for the DCS are compared with measurements carried out without laser assistance. The shift brought to the FF by the radiation is in the order of some percent. 

Coherent X-ray light with high intensity, which have become recently available due to FEL sources \cite{Emma2010, McN2010, Ish2012}, can be used jointly with the theoretical formalism presented in this Letter to explore quantized shifts in FFs of mesoscopic, atomic and also nuclear targets. Moreover, our results open up the possibility to measure FF values by varying the linear momentum of the ER ($\hbar\vec k$), for fixed electron-target linear momentum transfer ($\vec Q$). This allows, for example, to measure FFs at zero argument in scattering events for which the electron-target linear momentum transfer is {\it not} zero.
The elastic FF at zero argument, although it carries important information on the quadratic mean radius of spherically symmetric charge distributions \cite{Povh2006, Eur}, is normally difficult to measure in the absence of ER, since in that case the signal of scattering events is suppressed by the background of non-scattering events.

Finally, experiments on free electron scattering by ions have recently attracted much interest due to their straightforward applications to plasma and astrophysics \cite{Sri1996, Mul2008}. Understanding the ER effects on the pattern of the scattered electrons and on the target FFs might thus have direct impact on plasma diagnostics and astrophysics.

SI units are used throughout this letter. 

\bigskip

{\bf\raggedleft Electron state inside the radiation}

Let us consider linearly polarized ER whose vector potential can be taken of the form $\vec A=\vec A_0\sin(\vec k\cdot\vec r-\omega t)$. Here, $\omega$ is the angular frequency while $\vec k$ is the wave-vector of the ER ($ c |\vec k|=\omega$,  $c$ is the speed of light in vacuum). Within the Coulomb gauge and by retaining only terms linear in $\hbar\omega/(mc^2)$, the free electron state inside the radiation can be described by the wave-function \cite{MkA2009, Fe1993}
\begin{equation}
\label{eq:PsiFinal}
\begin{array}{l}
\ds
\Psi_{\vec P, \vec k}(\vec r, t) = 
\ds\frac{e^{\frac{i}{\hbar}(\vec P\cdot\vec r-E t)}}{\sqrt{V}}\,
\exp\left[i\alpha\big(1-\cos(\vec k\cdot\vec r - \omega t)\big)\right]\\[0.5cm]
\hspace{0.4cm}\ds\times\,\exp\left[
-i\beta\left( \vec k\cdot\vec r-\omega t-\frac{1}{2}\sin\big(2(\vec k\cdot \vec r -\omega t)\big) \right)
\right]~,
\end{array}
\end{equation}
where $V$ is the volume where the wave function is defined, $\alpha=e\vec P\cdot \vec A_0\big/\big(\hbar (\vec P\cdot \vec k-m\omega) \big)$, while $\beta=e^2\vec A_0^2\big/\big(4\hbar(\vec P\cdot\vec k-m\omega)\big)$, and $m$ and $e$ are the electron mass and charge, respectively. 
The continuous quantum numbers $\vec P$ and $E$ in \eqref{eq:PsiFinal} are conserved quantities used to label the electron state inside the ER. However, they do {\it not} represent the linear momentum and the energy of the electron (which, on the other hand, are not conserved quantities), as long as the ER is on. They merely identify the electron state inside the ER and are called \supercomas{linear momentum} and \supercomas{energy}
parameters \cite{MkA2009}. Nevertheless, if the ER is switched off, the wave function \eqref{eq:PsiFinal} will become a plane-wave and, consequently, $\vec P$ and $E$ will then correctly represent the conserved linear momentum and energy of the electron, respectively. 
In either cases the ER is on or off, the energy-momentum relation $2mE=\vec P^2$ must be satisfied \cite{Ber1982}.
Finally, we notice that the parameters $\alpha$ and $\beta$ contain the dependence on the angle between the linear momentum parameter ($\vec P$) and the ER wave-vector ($\vec k$).

\bigskip

{\bf\raggedleft Scattering differential cross section}

Let us now consider LAES by some target characterized by an
extended charge distribution and let us denote by $\theta$ the scattering angle.
The electron-target interaction is assumed to be electrostatic and the space where the scattering happens is assumed to be filled with linearly polarized ER. 
We suppose that the ER be non-invasive, i.e., that the target is not perturbed by the ER, as usual in LAES \cite{Mas1993}.
The amplitude for this scattering process can be written in first order time-dependent perturbation theory as \cite{Ber1982}
\begin{equation}
\label{eq:Ampl2}
\begin{split}
\mathcal{A} &= 
\ds (i\hbar)^{-1} e\int_{-t}^{t} \dd t' \int\dd^3 \vec r\,
\Psi^*_{\vec P_f, \vec k}(\vec r, t') \,V(\vec r) \, \Psi_{\vec P_i, \vec k}(\vec r, t')~,
\end{split}
\end{equation}
where $\Delta t=2t$ is the scattering time interval and the potential $V(\vec r)$ is of the form \cite{Povh2006, Bren2008, Messiah1962}
\begin{equation}
\label{eq:Vpot}
V(\vec r) = \int \dd^3\vec r_{\xi} \;\phi_f^*(\vec r_{\xi})\, 
V_p\big(|\vec r - \vec r_{\xi}|\big) \,
\phi_i(\vec r_{\xi})
~.
\end{equation}
Here $\phi_{i, f}$ are the initial and final target wave-functions, while $V_p\big(|\vec r - \vec r_{\xi}|\big)$ denotes the electrostatic potential between two point-like charges at distance $|\vec r - \vec r_{\xi}|$. We have here assumed, for simplicity, that the target can be described with a single variable $\vec r_{\xi}$: The extension to many-body targets is trivial. The amplitude (\ref{eq:Ampl2}) does not fully take into account the bound structure of the target. Rather, the target is treated as a source of potential. The second order scattering term should be included in order to fully account for the target bound structure and resonances. Consequently, the validity range of (\ref{eq:Ampl2}) is restricted to electron energies far from target resonances.

The scattering amplitude \eqref{eq:Ampl2} may be further manipulated by employing the Jacobi-Anger expansion of the exponentials. The evaluation of the amplitude $\mathcal{A}$ proceeds straightforwardly and is similar to the standard evaluation in the absence of ER, as showed, for example, in Refs. \cite{Bren2008, Povh2006}. Mathematical details are given in the supplementary material \cite{Suppl}.
On squaring the amplitude, extending the scattering time interval to infinity for obtaining energy conservation, multiplying by the density of final states and normalizing to the electron flux, we obtain the differential cross section for electron scattering in the presence of linearly polarized ER as
\begin{equation}
\begin{array}{lcl}
\label{eq:DiffCrossSecComplete}
\ds\frac{\dd \sigma_{s}}{\dd\Omega} (\vec P_i, \vec P_f)&=&\ds 
\frac{|\vec P_f|}{|\vec P_i|} \,J_s^{\,2}(\alpha_i-\alpha_f)\,
\left(\frac{2m \hbar\,c\, \alpha_e }{(\vec Q + s\hbar \vec k)^2}\right)^2\\[0.4cm]
&&\times\;\big| F(\vec Q + s\hbar \vec k) \big|^2 ~,
\end{array}
\end{equation}
where $J_s$ is the Bessel function of order $s$, $E_f=E_i+s\hbar\omega=\vec P_f^2/(2m)$ is the energy of the outgoing electron,
$\vec Q=\vec P_i - \vec P_f$ is, as mentioned above, the linear momentum transfer between electron and target, $s$ is an integer number ranging from $-\infty$ to $+\infty$ representing the photon number. In addition, $\alpha_{i(f)}$ is the previously defined quantity $\alpha$ calculated for $\vec P\to \vec P_{i(f)}$, while $\alpha_e$ is the electromagnetic coupling constant. 

The quantity $F(\vec Q)$ is the target FF and reads
\begin{equation}
\label{eq:FF}
F(\vec Q)=\ds\int\dd^3 \vec r\,e^{\frac{i}{\hbar}\vec r\cdot\vec Q}\,\phi_f^*(\vec r)\phi_i(\vec r)~.
\end{equation}
In elastic scattering, we have $\phi_i=\phi_f$, and, consequently, the FF represents the Fourier transform of the target charge distribution. In this case, the FF is referred to as elastic FF. On the other hand, if the scattering is inelastic, we have $\phi_i\neq\phi_f$, and the FF is referred to as inelastic FF \cite{Bren2008}.
In deriving Eq. \eqref{eq:DiffCrossSecComplete}, we have furthermore assumed $\beta_i-\beta_f\approx0$, which is justified within the non-relativistic assumption.
Here $\beta_{i(f)}$ is the previously defined quantity $\beta$ calculated for $\vec P\to \vec P_{i(f)}$.

In the case the target is an isolated atom, the scattering cross section gets contributions from both the nucleus and the atomic electrons, for which the relative form factors add coherently but with opposite sign. For non-relativistic energies, we may set the nuclear form factor equal to the nuclear charge ($Z_n$). Therefore, in this case Eq. \eqref{eq:DiffCrossSecComplete} reads
\begin{equation}
\label{eq:DiffCrossSecAt}
\begin{array}{lcl}
\ds\frac{\dd \sigma_{s}}{\dd\Omega} (\vec P_i, \vec P_f)&=&\ds 
\frac{|\vec P_f|}{|\vec P_i|} \,J_s^{\,2}(\alpha_i-\alpha_f)\,
\left(\frac{2m \hbar\,c\, \alpha_e }{(\vec Q + s\hbar \vec k)^2}\right)^2\\[0.4cm]
&&\times\;\big|Z_n - F_{\textrm{At}}(\vec Q+ s\hbar \vec k) \big|^2~,
\end{array}
\end{equation}
where $F_{\textrm{At}}(\vec Q)$ is the atomic form factor. If the ER is switched off (which is accomplished by setting $\alpha_{i,f}, \vec k\to0$), Eq. \eqref{eq:DiffCrossSecAt} turns out to be equal to the Mott formula \cite{Bren2008}, as expected.

Since Eq. \eqref{eq:DiffCrossSecComplete} has been derived in non-relativistic first order perturbation theory, it is not directly applicable when the electron energy is relativistic ($E_{i,f}\gtrsim 100$ keV) or when the scattering angle is particularly small ($\theta\lesssim 2^\circ$). However, relativistic, screening and spin corrections can be added in a similar way as done for the scattering cross section in the absence of ER, so as to make it applicable for a wider energy range and for small scattering angles \cite{Wil2009, Povh2006}. 

By analyzing Eq. \eqref{eq:DiffCrossSecComplete}, we notice that the energies of the scattered electrons turn out to be spanned by multiples of the photons energy. The integer number $s$ is thus easily interpreted as the number of photons absorbed (if $s>0$) or emitted (if $s<0$) by the electron during the scattering process, as normally done in LAES experiments \cite{deHL2011}.
Most importantly, we notice that the presence of ER shifts the argument of the FF from $\vec Q$ to $\vec Q+s\hbar\vec k$. The photon number $s$, on which the shift depends, can be measured by detecting the kinetic energy of the scattered electron. Due to the fact that such shift does not depend on the ER polarization, it will hold for any kind of ER polarization and, also, for unpolarized light. 

It must be underlined that Eq. \eqref{eq:DiffCrossSecComplete} is different from Eq. (210) of Ref. \cite{El1998}, as our $\bm Q$ corresponds to $\bm Q_N$ in Ref. \cite{El1998}.

\begin{figure}[t]
\includegraphics[scale=0.35]{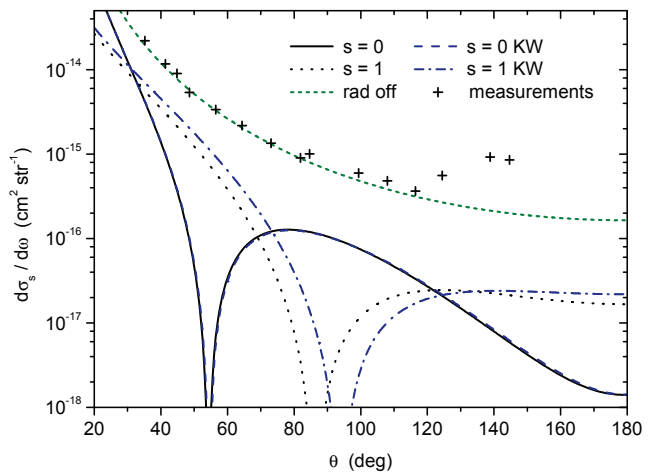}
\caption{(color online). Differential cross section for elastic electron scattering by Ne-like Argon in the presence of intense ultraviolet ER with wavelength $\simeq $ 224 nm. The electron initial energy is $E_i=22.46$ eV. The ER has: direction orthogonal to the scattering axis, linear polarization parallel to the linear momentum transfer, and intensity $I\simeq 10^{15}$  W/cm$^2$. The green short-dashed line refers to electron scattering in absence of ER, while the crosses represent the best values from experimental measurements taken in absence of ER \cite{Bel1996}. The curves denoted by KW are obtained with KWF. $s$ represents the number of photons absorbed (if $s>0$) or emitted (if $s<0$) by the scattered electron.
}
\label{fig:fig1}
\end{figure}

\bigskip

{\bf\raggedleft Kroll-Watson formula re-visited}

The Kroll-Watson formula (KWF) can be obtained either within the Born approximation, which is in line with our derivation, or within the low frequency limit ($\hbar\omega\ll e\bm P\cdot \bm A_0/m$) \cite{KrW1973}. In the light of this, we expect to recover the KWF as a special case of our theory, in the case of elastic scattering, within the assumptions of the first Born approximation.
Indeed, if the momentum carried by the absorbed or emitted photons is much lower than the momentum transfer between
electron and target, we may use $\vec Q + s\hbar\vec k\approx \vec Q$. 
Furthermore, the nonrelativistic assumption allows for the replacement $\vec P_{i,f}\cdot \vec k-m\omega\approx -m\omega$ in the denominators of $\alpha_{i,f}$.
Employing these two approximations in \eqref{eq:DiffCrossSecComplete} yields the KWF for the differential cross section
\begin{equation}
\label{eq:DiffCrossSecKWA}
\frac{\dd \sigma_{s}}{\dd\Omega}(\vec P_i, \vec P_f) \simeq
\frac{|\vec P_f|}{|\vec P_i|} \,J_s^{\,2}(\eta)\,\frac{\dd\sigma_{\el}}{\dd\Omega}(\vec Q) ~,
\end{equation}
where $\eta=-e\vec Q\cdot\vec A_0/(\hbar\omega m)$. Kroll and Watson obtained Eq. \eqref{eq:DiffCrossSecKWA}
by analyzing the Green function that governs the LAES process.
Our result \eqref{eq:DiffCrossSecComplete} can be thus considered a refinement of the KWF which allows to grasp the effect that the ER brings to the target FF. 

The KWF has been widely investigated in atomic physics and has been found to adequately describe the data in several experimental cases (see Ref. \cite{deHL2011} for a brief account). 
However, it has been showed that the KWF is unable to describe LAES when the scattering angle is small ($\theta\simeq 9^\circ$) \cite{WaH1994a} and when the laser polarization is orthogonal to the linear momentum transfer ($\vec A_0\perp \vec Q$) \cite{MuM2010}. 
Although a few attemps to solve the discrepancies have been made \cite{RaM1994, MaK1995, CiD1997, SuZ1997}, the problem is still open \cite{SiB2011}. 
On account of this, we ought to say that the validity of \eqref{eq:DiffCrossSecComplete} may share the same limitations of KWF.

\bigskip

\begin{figure}[t]
\includegraphics[scale=0.35]{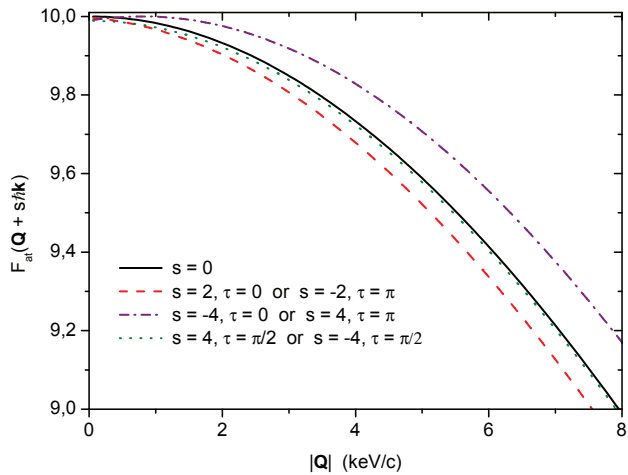}
\caption{(color online). Atomic form factor of Ne-like Argon in the presence of ER with wavelength $\simeq $ 6.2 nm. The parameter $\tau$ represents the angle between the electron-target linear momentum transfer and the ER direction, i.e. $\cos\tau=\vec k\cdot \vec Q/(|\vec k||\vec Q|)$. For the parameter $s$, see Fig. \ref{fig:fig1}.
}
\label{fig:fig2}
\end{figure}

{\bf\raggedleft Atomic examples}

Equation \eqref{eq:DiffCrossSecComplete} is the main result of this letter and describes the differential cross section for LAES.
Specifically, we read from Eq. \eqref{eq:DiffCrossSecComplete} that, in the presence of ER, the target FFs and the differential cross section depend not only on the momentum transfer $\vec Q$, but also on the ER wavelength and direction, as well as on the number of photons absorbed or emitted during the scattering process. 
Thus, these three last parameters can be suitably tuned, in experiments, so as to alter the shape of the target FF to be measured, for a given $\vec Q$. 
In the following, we discuss two examples to better highlight the impact of ER on the DCS and on the target FFs.

Let us consider elastic electron scattering by an isolated Ne-like Argon ion, in the presence of intense ultraviolet ER of wavelength $\simeq $ 224 nm (the wavelength of HeAg lasers). In this energy range, the ER may not excite the target, since its first excitation line is at $\sim$ 4.95 nm. The target is thus unaffected by the ER, as hypothesized. 
In Fig. \ref{fig:fig1}, we show the differential cross section, $\dd\sigma_s/\dd\Omega$, for such a process, where electron energy and ER parameters are specified. The experimental data agree well with the theoretical predictions for angles $\theta\lesssim120^\circ$. The theoretical predictions are given by Eq. \eqref{eq:DiffCrossSecAt} with $\alpha_{i,f}, \vec k\to0$, which equals the Mott formula. This also demonstrates that the free Coulomb wave-function assumed in Eq. \eqref{eq:PsiFinal} is suitable for describing scattering of electrons by Ne-like Argon as provided in the example. For larger angles, there are discrepancies caused by the fact that the slow scattering electron probes the bound structure of the ion, which is not taken into account by the Mott formula, as underlined in introducing Eq. \eqref{eq:Ampl2}. Nevertheless, such discrepancies can be removed if Hartree-Fock calculations are used \cite{Bel1996}. 

In Fig. \ref{fig:fig2}, we show the atomic FF of Ne-like Argon target in the presence of soft X-ray ER of wavelength $\simeq$ 6.2 nm, which is still higher than the first single photon absorption line. The atomic FF of Ne-like Argon has been analytically calculated using the novel technique presented in Ref. \cite{Lal2013}. As we see from Fig. \ref{fig:fig2}, the shift brought to the FF by the ER is in the order of few percent, and it depends on the two variables $\tau$ and $s$. 
We remark that, in order to have a probability ratio $\dd\sigma_{s\neq0}/\dd\sigma_{s=0}\sim1$ or greater with soft X-ray ER, radiation intensities as high as I $\gtrsim$ 10$^{18}$ W/cm$^2$ must be employed, which might generate two-photon absorption peaks. Even so, since the first ionization threshold of Ne-like Ar is at $\sim 2.94$ nm, two-photon absorption may not cause ionization of the target. 
Such intensities in the soft-X ray regime are nowadays achievable by using XFEL sources \cite{Emma2010, McN2010, Ish2012}.

\begin{figure}[b]
\includegraphics[scale=0.59]{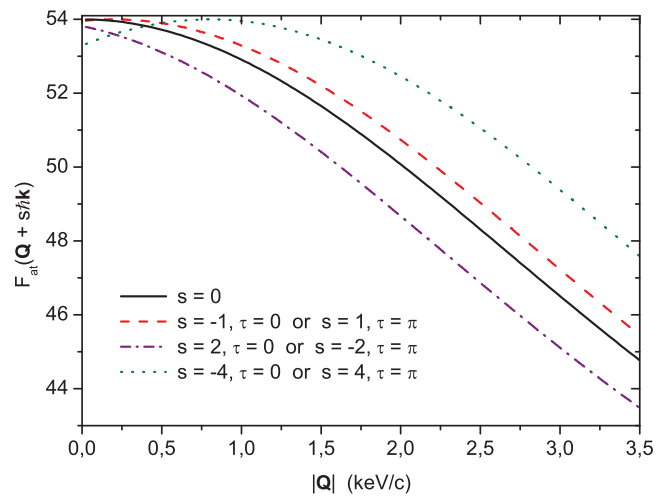}
\caption{(color online). Atomic form factor of Xe atom in the presence of ER with wavelength $\simeq $ 6.2 nm. See Figs. \ref{fig:fig2} and \ref{fig:fig1} for the parameters $\tau$ and $s$.
}
\label{fig:fig3}
\end{figure}

\medskip

The investigation of the linear momentum shift in the FF of neutral samples can be also done, although same care is needed.
In order to obtain a considerable shift in neutral atoms, one needs high intensity radiation whose photons have energy higher than few hundreds electonvolts. Under these conditions neutral samples will undergo several ionizations, leading to numerous background events. However, control parameters can be used to monitor the sample damage, as done in crystallography for monitoring the crystal damage \cite{Giac1}. Alternatively, one may use FEL sources and molecular structures. In fact, it has been proved very recently that FEL prevents sample damage in protein structures by simply outrunning it \cite{Chap1}. In view of these considerations, we present in Fig. \ref{fig:fig3} the shift brought to the FF of neutral Xe by soft X-ray ER of wavelength $\simeq$ 6.2 nm (same as in Fig. \ref{fig:fig2}). The FF shift is larger than in Fig. \ref{fig:fig2}, although still in the order of some percent. The choice of Xe atom is motivated by a recent LAES experiment carried out using $fs$ pulses with wavelength of about $800$ nm, intensity $10^{12}$ W/cm$^2$ \cite{Yama2010}.

The problem of sample damage on neutral sample might be also circumvented by further increasing the ER intensity so as to reach a new stability regime (stability given by super-intense laser pulses) \cite{Ebe1993, Eichmann}. 

\bigskip

{\bf\raggedleft Remarks}

Linear momentum shifts in form factors of clusters and mesoscopic objects can also be studied with the very same formalism developed here \cite{Ger1997, Col2003, Puddu, Heer}. A FF shift of the order $\sim $ eV/c is already significant in such systems, due to their larger size. This can be achieved by using slow electrons and visible light.

As for the case of the KWF \cite{Kam1985}, the present formalism may be extended to a relativistic framework by employing Volkov solutions and the full electron-target electromagnetic interaction \cite{Ber1982}.
Such an extension should permit to measure electric and magnetic FFs values at zero argument from scattering events whose momentum transfer is not zero. 
This would help solve the puzzle arisen from recent measurements on the proton structure, where it has been showed that the ratio between electric and magnetic FFs, at $\vec Q\approx 0$, is slightly less than what is to be expected from QCD considerations \cite{Ro2007, Zh2011}. The presented formalism might be also employed to shed light onto recent disputed measurements of the proton radius \cite{Eur, Pol}.

Finally, Form Factor shifts can be also studied with both radiation-free \cite{Mat} and radiation-assisted \cite{Arm} twisted electrons. Such studies will be addressed in future publications.

\begin{widetext}
\begin{center} \bf Supplementary material: Mathematical details \end{center}

The amplitude for the scattering process in Eq. (2) of the text can be written in full glory as 
\begin{equation}
\label{eq:A1}
\begin{array}{lcl}
\mathcal{A}
&=&\ds\frac{(-i\hbar)^{-1}}{V}\int_{-t}^{t}\dd t'\int\dd^3\vec r\left[
e^{-\frac{i}{\hbar}t'(E_i-E_f)}e^{\frac{i}{\hbar}\vec r(P_i-P_f)}e^{-i(\alpha_i-\alpha_f)}
e^{-i(\alpha_i-\alpha_f)\cos\left[\vec k \vec r-\omega t'\right]}
e^{-i\vec k \vec r(\beta_i-\beta_f)}\right.\\
&&\qquad\qquad\times\,\ds\left.e^{i\omega t'(\beta_i-\beta_f)}
e^{i\frac{\beta_i-\beta_f}{2}\sin\left[2(\vec k\vec r-\omega t')\right]}\,V(r)
\right]~.
\end{array}
\end{equation}
On the equation above we shall use the Jacobi-Anger expansions.
After having integrated over $\dd t'$, from Eq. \eqref{eq:A1} we find
\begin{equation}
\label{eq:A2}
\begin{array}{lcl}
\mathcal{A}&=&\ds\frac{e^{i(\alpha_i-\alpha_f)}}{V}\sum_{s,s_1}
J_s(\alpha_i-\alpha_f)J_{s_1}\left(\frac{\beta_i-\beta_f}{2}\right)
\frac{e^{-is\frac{\pi}{2}}}{\epsilon}\left(-2i\sin\left[\frac{t\epsilon}{\hbar}\right]\right)
\int\dd^3 \vec r\,e^{\frac{i}{\hbar}\vec r\vec \chi}\, V(r)~,
\end{array}
\end{equation}
where $\epsilon=E_i-E_f+\hbar\omega(\beta_f-\beta_i+s+2s_1)$ and $\vec \chi= \vec P_i-\vec P_f+\hbar\vec k(\beta_f-\beta_i+s+2s_1)$.
We rewrite the integral above by making use of the equivalence
\begin{equation}
\begin{array}{lcl}
\label{eq:Poisson}
\ds\int\dd^3 \vec r \,e^{\frac{i}{\hbar}\vec r\vec \chi}\,V(r)
&=&\ds\int\dd^3 \vec r \left(-\frac{\hbar^2}{\chi^2}\right)\,\big(\nabla^2 e^{\frac{i}{\hbar}\vec r\vec \chi}\big)\,V(r)
=-\frac{\hbar^2}{\chi^2}\int\dd^3 \vec r \,e^{\frac{i}{\hbar}\vec r\vec \chi}\,\big(\nabla^2 V(r)\big)\\[0.4cm]
&=&\ds\left(-\frac{4\pi\hbar^3 c\alpha_e }{\chi^2}\right)\int\dd^3 \vec r e^{\frac{i}{\hbar}\vec r\vec \chi} 
\big( \phi_f^*(\vec r)\phi_i(\vec r) \big)
\end{array}
\end{equation}
where we used $\nabla^2\frac{1}{|\vec r-\vec r'|}=4\pi\delta^3(\vec r-\vec r')$, and where $\alpha_e$ is the electromagnetic coupling constant. In writing Eq. \eqref{eq:Poisson}, we have discarded the surface term coming from the partial integration, since the potential $V(r)$ and its derivative are vanishing for $r\to +\infty$.
Joining Eq. \eqref{eq:Poisson} with Eq. \eqref{eq:A2}, we readily get
\begin{equation}
\label{eq:A3}
\begin{array}{lcl}
\mathcal{A}&=&\ds\left(+8i\pi\hbar^3 c\alpha_e \right)
\frac{e^{i(\alpha_i-\alpha_f)}}{V}\sum_{s,s_1}
J_s(\alpha_i-\alpha_f)J_{s_1}\left(\frac{\beta_i-\beta_f}{2}\right)
\frac{e^{-is\frac{\pi}{2}}}{\epsilon\chi^2}\sin\left[\frac{t\epsilon}{\hbar}\right]
F(\vec \chi)~,
\end{array}
\end{equation}
where $F(\vec \chi)=\ds\int\dd^3 \vec r e^{\frac{i}{\hbar}\vec r\vec \chi} \big( \phi_f^*(\vec r)\phi_i(\vec r) \big)$.
In order to obtain the differential probability for the process ($\dd O$), we take the modulus squared of the amplitude and multiply it by the density of final states: $\dd O\equiv|\mathcal{A}|^2\frac{V\dd^3 P_f}{(2\pi\hbar)^3}$. 
The scattering rate ($\dd W$) is then obtained by taking the first derivative of $\dd O$ with respect to the scattering time interval $\Delta t=2t$, and by sending $\Delta t\to +\infty$, for obtaining energy conservation. Consequently, the oscillatory terms with different frequencies give zero contribution and after a few algebraic passages we are left with
\begin{equation}
\label{eq:A4}
\begin{array}{lcl}
\ds \dd W
&=&\ds \frac{4\hbar^2c^2\alpha_e^2\sqrt{2E_f}m^{3/2}}{V}
\sum_{s,s_1}\left(J_s(\alpha_i-\alpha_f)J_{s_1}\left(\frac{\beta_i-\beta_f}{2}\right)\right)^2
\frac{|F(\vec \chi)|^2}{\chi^4}\;
\delta(\epsilon) \;\dd E_f \dd \Omega_{P_f}
\end{array}
\end{equation}
where we used  
$
\lim_{\Delta t\to +\infty} \frac{\sin\left[\frac{\epsilon\Delta t}{\hbar}\right]}{\epsilon/\hbar}
=\pi \delta\big(\epsilon/\hbar\big)=\pi \hbar\delta\big(\epsilon\big)~
$ and $\dd^3 P_f= m^{3/2}\sqrt{2E_f}\dd E_f \dd \Omega_{P_f}$. Here, $\dd \Omega_{P_f}$ is the angle differential of the vector $\vec P_f$.
In order to obtain the differential cross section ($\dd \sigma$), we must normalize the scattering rate to the initial electron flux \cite{Povh2006}.
The initial electron flux can be written as $J=<v^a>/V$, where $<v^a>$ is the average initial velocity of the electron beam along the scattering axis (i.e., the axis between the electron at initial time and the target). The quantity $<v^a>$ can be calculated from
\begin{equation}
\label{eq:vv}
<v^a>=\frac{1}{m}\int\dd^3 \vec r\,\Psi_{\vec P_i, \vec k}^*(\vec r)\opscalar p^a\Psi_{\vec P_i, \vec k}(\vec r)
=\frac{1}{m}( P_i^a-\hbar \beta_i k^a)~,
\end{equation}
where $\opscalar p^a$ is the linear momentum operator along the scattering axis, and $P_i^a$, $k^a$ are the component along the scattering axis of the electron initial linear momentum and of the radiation wave-vector, respectively. For moderate laser intensities (or by assuming the radiation direction to be orthogonal to the scattering axis), we can approximate $P_i^a-\hbar \beta_i k^a\approx P_i^a=|\vec P_i|$, where the last step follows from the fact that the electron initial linear momentum is normally along the scattering axis. With the help of Eq. \eqref{eq:vv} and by approximating $\beta_i-\beta_f\approx 0$, we may finally write 
\begin{equation}
\label{eq:A5}
\begin{array}{lcl}
\ds \frac{\dd \sigma}{\dd E_f\dd \Omega_{P_f}}&=&\ds\frac{1}{J}\frac{\dd W}{\dd E_f\dd \Omega_{P_f}}=4\hbar^2c^2\alpha_e^2m^2\frac{|\vec P_f|}{|\vec P_i|}
\;
\sum_{s=-\infty}^{+\infty}
J_s^2(\alpha_i-\alpha_f)\frac{|F(\vec Q+s\hbar\vec k)|^2}{(\vec Q+s\hbar\vec k)^4}
\;\delta\big(E_i-E_f+\hbar\omega s\big)~,
\end{array}
\end{equation}
where $\vec Q=\bm P_i-\bm P_f$ is the momentum transferred in the scattering process. By integrating over the energy of the scattered electron and by selecting the parameter $s$ \cite{footnote}, we find Eq. (4) of the text.
\end{widetext}

\begin{acknowledgments}
F.F. acknowledges support by i) Funda\c{c}\~ao de Amparo \`a Pesquisa do estado de Minas Gerais (FAPEMIG) and ii) Conselho Nacional de Desenvolvimento Cient\'ifico e Tecnol\'ogico (CNPq).
L.S., K.J. and F.F. acknowledge support by the Research Council for Natural Sciences and Engineering of the Academy of Finland.
A.G.H. acknowledges the support from the GSI Helmholtzzentrum and the University of Heidelberg.
P. A. acknowledges support by the German Research Foundation (DFG) within the Emmy Noether program under 
Contract No. TA 740 1-1. 
J. P. S. and P. A. acknowledge support by FCT -- Funda\c{c}\~ao para a Ci\^encia e a Tecnologia (Portugal), 
through the Projects No. PEstOE/FIS/UI0303/2011 and PTDC/FIS/117606/2010, financed by the European Community 
Fund FEDER through the COMPETE -- Competitiveness Factors Operational Programme. \\
F.F. is thankful to Prof. Marcelo Fran\c{c}a Santos, Prof. Sergio Scopetta, Dr. Wei Cao and Dr. Abraham Kano for useful discussions and comments.
\end{acknowledgments}



\begin{thebibliography}{13}
\bibitem{footnote}				This can be done by monitoring the kinetic energy of the scattered electron.
\end{thebibliography}
\end{document}